 
\input harvmac
\input amssym

\def\coeff#1#2{\relax{\textstyle {#1 \over #2}}\displaystyle}

\def\oneone{\rlap 1\mkern4mu{\rm l}}
\def\eql{~=~}

\def\cA{{\cal A}}

\def\cH{{\cal H}} 
 
\def\cL{{\cal L}} \def\cM{{\cal M}}
\def\cN{{\cal N}} \def\cO{{\cal O}}
 
\def\cR{{\cal R}} \def\cS{{\cal S}}

\def\bfone{\relax{\rm 1\kern-.35em 1}}
\def\IC{\relax\,\hbox{$\inbar\kern-.3em{\rm C}$}}
\def\ID{\relax{\rm I\kern-.18em D}}
\def\IF{\relax{\rm I\kern-.18em F}}
\def\IH{\relax{\rm I\kern-.18em H}}
\def\II{\relax{\rm I\kern-.17em I}}
\def\IN{\relax{\rm I\kern-.18em N}}
\def\IP{\relax{\rm I\kern-.18em P}}
\def\IQ{\relax\,\hbox{$\inbar\kern-.3em{\rm Q}$}}

\def\IR{\relax{\rm I\kern-.18em R}}
\font\cmss=cmss10 \font\cmsss=cmss10 at 7pt
\def\ZZ{\relax\ifmmode\mathchoice
{\hbox{\cmss Z\kern-.4em Z}}{\hbox{\cmss Z\kern-.4em Z}}
{\lower.9pt\hbox{\cmsss Z\kern-.4em Z}}
{\lower1.2pt\hbox{\cmsss Z\kern-.4em Z}}\else{\cmss Z\kern-.4em
Z}\fi}

\def\ZZ{\Bbb{Z}} 
\def\IC{\Bbb{C}}
\def\ID{\Bbb{D}}
\def\IF{\Bbb{F}}
\def\IH{\Bbb{H}}
\def\II{\Bbb{I}}
\def\IN{\Bbb{N}}
\def\IP{\Bbb{P}}
\def\IQ{\Bbb{Q}}
\def\IR{\Bbb{R}}
%
%
 %
 %
\lref\LeighEP{
R.~G.~Leigh and M.~J.~Strassler,
``Exactly marginal operators and duality in four-dimensional N=1 supersymmetric
gauge theory,''
Nucl.\ Phys.\ B {\bf 447}, 95 (1995)
[arXiv:hep-th/9503121].
}
%
\lref\FreedmanGP{
D.~Z.~Freedman, S.~S.~Gubser, K.~Pilch and N.~P.~Warner,
``Renormalization group flows from holography supersymmetry and a  c-theorem,''
Adv.\ Theor.\ Math.\ Phys.\  {\bf 3}, 363 (1999)
[arXiv:hep-th/9904017].
}
%
\lref\PilchEJ{
K.~Pilch and N.~P.~Warner,
``A new supersymmetric compactification of chiral IIB supergravity,''
Phys.\ Lett.\ B {\bf 487} (2000) 22,
hep-th/0002192.
}
%
\lref\PilchFU{
K.~Pilch and N.~P.~Warner,
``N = 1 supersymmetric renormalization group flows from IIB supergravity,''
Adv.\ Theor.\ Math.\ Phys.\  {\bf 4} (2002) 627,
hep-th/0006066.
}
%
\lref\KhavaevFB{
A.~Khavaev, K.~Pilch and N.~P.~Warner,
``New vacua of gauged N = 8 supergravity in five dimensions,''
Phys.\ Lett.\ B {\bf 487} (2000) 14,
hep-th/9812035.
}
%
\lref\KhavaevGB{
A.~Khavaev and N.~P.~Warner,
 ``A class of N = 1 supersymmetric RG flows from 
 five-dimensional N = 8 supergravity,''
Phys.\ Lett.\ B {\bf 495}  (2000) 215,
 hep-th/0009159.
}
%
\lref\KhavaevYG{
A.~Khavaev and N.~P.~Warner,
``An N = 1 supersymmetric Coulomb flow in IIB supergravity,''
Phys.\ Lett.\ B {\bf 522} (2001) 181,
hep-th/0106032.
}
%
\lref\JohnsonIC{
C.~V.~Johnson, K.~J.~Lovis and D.~C.~Page,
``Probing some N = 1 AdS/CFT RG flows,''
JHEP {\bf 0105} (2001) 036,
hep-th/0011166.
}
%
\lref\JohnsonZE{
C.~V.~Johnson, K.~J.~Lovis and D.~C.~Page,
``The K\"ahler structure of supersymmetric holographic RG flows,''
JHEP {\bf 0110} (2001) 014,
hep-th/0107261.
}
%
\lref\SchwarzQR{
J.~H.~Schwarz,
``Covariant Field Equations Of Chiral N=2 D = 10 Supergravity,''
Nucl.\ Phys.\ B {\bf 226}, 269 (1983).
}
\lref\HoweSR{
  P.~S.~Howe and P.~C.~West,
``The Complete N=2, D = 10 Supergravity,''
  Nucl.\ Phys.\ B {\bf 238}, 181 (1984).
}
%
\lref\SchwarzWA{
  J.~H.~Schwarz and P.~C.~West,
``Symmetries And Transformations Of Chiral N=2 D = 10 Supergravity,''
  Phys.\ Lett.\ B {\bf 126}, 301 (1983).
}
%
%
\lref\HalmagyiJY{
N.~Halmagyi, K.~Pilch, C.~Romelsberger and N.~P.~Warner,
``The complex geometry of holographic flows of quiver gauge theories,''
arXiv:hep-th/0406147.
}
%
\lref\PilchYG{
K.~Pilch and N.~P.~Warner,
``N = 1 supersymmetric solutions of IIB supergravity from Killing spinors,''
arXiv:hep-th/0403005.
}
\lref\CorradoNV{
R.~Corrado, K.~Pilch and N.~P.~Warner,
``An N = 2 supersymmetric membrane flow,''
Nucl.\ Phys.\ B {\bf 629} (2002) 74,
hep-th/0107220.
}
%
\lref\GiddingsYU{
  S.~B.~Giddings, S.~Kachru and J.~Polchinski,
``Hierarchies from fluxes in string compactifications,''
  Phys.\ Rev.\ D {\bf 66}, 106006 (2002)
  [arXiv:hep-th/0105097].
}
%
\lref\GranaXN{
  M.~Grana and J.~Polchinski,
``Gauge / gravity duals with holomorphic dilaton,''
  Phys.\ Rev.\ D {\bf 65}, 126005 (2002)
  [arXiv:hep-th/0106014].
}
%
\lref\MyersPS{
  R.~C.~Myers,
``Dielectric-branes,''
  JHEP {\bf 9912}, 022 (1999)
  [arXiv:hep-th/9910053].
}
%
%
\lref\PopeJP{
 C.~N.~Pope and N.~P.~Warner,
``A dielectric flow solution with maximal supersymmetry,''
  JHEP {\bf 0404}, 011 (2004)
  [arXiv:hep-th/0304132].
}
%
\lref\GowdigereJF{
  C.~N.~Gowdigere, D.~Nemeschansky and N.~P.~Warner,
``Supersymmetric solutions with fluxes from algebraic Killing spinors,''
  Adv.\ Theor.\ Math.\ Phys.\  {\bf 7}, 787 (2004)
  [arXiv:hep-th/0306097].
}
%
\lref\PilchJG{
  K.~Pilch and N.~P.~Warner,
``Generalizing the N = 2 supersymmetric RG flow solution of IIB supergravity,''
  Nucl.\ Phys.\ B {\bf 675}, 99 (2003)
  [arXiv:hep-th/0306098].
}
%
\lref\NemeschanskyYH{
  D.~Nemeschansky and N.~P.~Warner,
``A family of M-theory flows with four supersymmetries,''
  arXiv:hep-th/0403006.
}
%
\lref\BenaJW{
  I.~Bena and N.~P.~Warner,
``A harmonic family of dielectric flow solutions with maximal supersymmetry,''
  JHEP {\bf 0412}, 021 (2004)
  [arXiv:hep-th/0406145].
}
%
\lref\GranaBG{
  M.~Grana, R.~Minasian, M.~Petrini and A.~Tomasiello,
``Supersymmetric backgrounds from generalized Calabi-Yau manifolds,''
  JHEP {\bf 0408}, 046 (2004)
  [arXiv:hep-th/0406137].
}
%
\lref\GranaSV{
  M.~Grana, R.~Minasian, M.~Petrini and A.~Tomasiello,
``Type II strings and generalized Calabi-Yau manifolds,''
  Comptes Rendus Physique {\bf 5}, 979 (2004)
  [arXiv:hep-th/0409176].
}
%
\lref\BeckerYV{
K.~Becker, M.~Becker, K.~Dasgupta and P.~S.~Green, I,''
JHEP {\bf 0304}, 007 (2003)
[arXiv:hep-th/0301161].
}
%
\lref\BeckerSH{
  K.~Becker, M.~Becker, P.~S.~Green, K.~Dasgupta and E.~Sharpe,
``Compactifications of heterotic strings on non-Kaehler complex  manifolds. II,''
  Nucl.\ Phys.\ B {\bf 678}, 19 (2004)
  [arXiv:hep-th/0310058].
}
%
\lref\CorradoWX{
  R.~Corrado, M.~Gunaydin, N.~P.~Warner and M.~Zagermann,
``Orbifolds and flows from gauged supergravity,''
  Phys.\ Rev.\ D {\bf 65}, 125024 (2002)
  [arXiv:hep-th/0203057].
}
%
\lref\KlebanovHH{
  I.~R.~Klebanov and E.~Witten,
``Superconformal field theory on threebranes at a Calabi-Yau  singularity,''
  Nucl.\ Phys.\ B {\bf 536}, 199 (1998)
  [arXiv:hep-th/9807080].
}
%



\Title{
\vbox{
\hbox{\tt hep-th/0505019}
}}
{\vbox{\vskip -1.0cm
\centerline{\hbox{Holographic Coulomb Branch Flows}}
\vskip 8 pt
\centerline{\hbox{with $\cN=1$ Supersymmetry}}}}
\vskip -.3cm
\centerline{Chethan N. Gowdigere and Nicholas P. Warner }
\medskip
\bigskip
\centerline{{${}^{(1)}$\it Department of Physics and Astronomy}}
\centerline{{\it University of Southern California}}
\centerline{{\it Los Angeles, CA 90089-0484, USA}}
\medskip

\bigskip
\bigskip
We obtain a large, new class of $\cN=1$ supersymmetric holographic flow
backgrounds with $U(1)^3$ symmetry.  These solutions correspond
to flows toward the Coulomb branch of the non-trivial $\cN=1$ supersymmetric 
fixed point.  The massless (complex) chiral fields are allowed to develop
vevs that are independent of their two phase angles, and this corresponds
to allowing the brane to spread with arbitrary,  $U(1)^2$ invariant, 
radial distributions in each of these directions.  Our solutions are ``almost
Calabi-Yau:''   The metric is hermitian with respect to an integrable complex
structure, but is not K\"ahler.  The ``modulus squared'' of the holomorphic 
$(3,0)$-form is the volume form, and the complete solution is characterized
by a function that must satisfy a single partial differential equation that
is closely related to the Calabi-Yau condition.  The deformation from a standard
Calabi-Yau background is driven by a non-trivial, non-normalizable $3$-form flux dual
to a fermion mass that reduces the supersymmetry to $\cN=1$. This flux also
induces dielectric polarization of the $D3$-branes into $D5$-branes.

\vskip .3in
\Date{\sl {May, 2005}}

\vfill\eject

\newsec{Introduction}
 
In this paper we continue our study of supersymmetric backgrounds in
string theory, and most particularly for holography.
In the more traditional compactifications of string theory the internal, or
compactifying manifold is either compact, or effectively compact\foot{
By effectively compact we mean non-compact, but with normalizable
background fields.}.  This makes the task of classifying supersymmetric backgrounds
somewhat  easier.  Indeed, for smooth solutions to IIB supergravity it has been shown 
 that  $\cN=1$ supersymmetry in four dimensions places some very stringent
constraints on the background fields.  For example \refs{\GiddingsYU, \GranaXN}, 
if there is no  dilaton then the compactification has to be a warped Calabi-Yau manifold 
with  an imaginary self-dual $3$-form flux.    However, the proof makes strong use
of the square-integrability of the background fields, and is therefore invalid for
non-compact or singular backgrounds with non-normalizable fields.  This 
exception is precisely what one wishes to study in holography, where 
the non-normalizable modes correspond to perturbations of the Lagrangian
of the holographic field theory.  Thus the study of supersymmetric, 
holographic flows is precisely a study in the exceptions to the theorem  of
\GiddingsYU.

There are now very large families of physically interesting non-compact
backgrounds with reduced supersymmetry.  Many of them are based upon
Calabi-Yau backgrounds, but some of them are complex, but not  K\"ahler 
\refs{\BeckerYV\BeckerSH\GranaBG\GranaSV{--}\HalmagyiJY}.    
The family we wish to focus on here is one of the simplest holographic flows:
One starts with the $\cN=4$ supersymmetric Yang-Mills theory and gives
a mass to a single $\cN=1$  chiral multiplet.  As is well-known, this perturbed theory
preserves $\cN=1$ supersymmetry and has a non-trivial infra-red fixed point \LeighEP.
The holographic description of this fixed point, and the flow to it, is also well 
studied \refs{\FreedmanGP\PilchEJ\PilchFU\HalmagyiJY\JohnsonIC{--}\JohnsonZE}.  
Indeed, there has 
been recent progress in understanding the underlying geometry in terms of 
spaces that are almost Calabi-Yau manifolds \HalmagyiJY.  
This paper further develops this work.   The two chiral multiplets that are not
given a mass remain massless along the flow and so the complete $\cN=1$
supersymmetric field theory also has a two complex-dimensional Coulomb 
branch.    A three-parameter family of flows on this Coulomb branch were
studied in \refs{\KhavaevGB,\KhavaevYG}.  One of the parameters was the mass of the 
chiral multiplet,  $\Phi_3$,  while the other two were independent vevs of $\Phi_1$ and 
$\Phi_2$. Since these solutions were based upon gauged supergravity, the vevs of
these fields were very restricted and corresponded to brane distributions that
spread uniformly in each of these directions.   Our purpose here is to analyze 
solutions in which the branes are allowed to spread with arbitrary radial distributions
 in each of these two directions.   This means the solutions will
depend upon three variables that correspond to the magnitudes, $\Phi_j$.
As in \HalmagyiJY,  we will be able to characterize our solutions in terms of a deformation
of the Calabi-Yau condition.

On a more technical level, we will proceed in the same spirit as 
\refs{\PopeJP\GowdigereJF\PilchJG\PilchYG\NemeschanskyYH{--}\BenaJW}, and 
use algebraic Killing spinors.  In the past, such calculations have
involved imposing a high level of symmetry so that the metric functions
and fluxes can {\it only} depend upon two variables.   In \PilchYG\ this led to a 
result that appeared to depend upon having only two-variables:  The
solution was determined by a single function $\Psi(u,v)$, but one also 
needed to construct a conjugate function, $S(u,v)$, that looked like
a non-linear analog of the harmonic conjugate of $\Psi(u,v)$.  It was thus
not clear whether the relatively simple results of  \PilchYG\ were an artifact of
the high level of symmetry.  In this paper we consider generalizations
of the flow of \PilchYG\ in which there is less symmetry, and the underlying functions
depend upon three variables.  We will show that the simplicity of the
result of \refs{\PilchYG,\HalmagyiJY} persists:  The non-trivial flow solutions arise 
from a deformation of the Calabi-Yau condition.  Indeed we found that deriving the related 
Calabi-Yau metric first provided remarkable insights into how to solve the
more general problem with non-trivial fluxes considered here.  In the process of finding the 
more general class of solutions we will also simplify and unify the results of 
\refs{\PilchYG,\HalmagyiJY}.

In section 2 we will briefly summarize the relevant field theory and use its
symmetries to constrain the Ansatz for the holographic theory.  In section 3 
we will make the complete Ansatz for the holographic background.  In section 4
we will find the ``wrong solution'' in that we will set the fluxes to zero and 
find the most general Calabi-Yau metric.  In section 5 we present the new solutions 
by showing how the Calabi-Yau equations are successively modified.  We
then show that the new solutions are ``almost Calab-Yau'' in that they have
an integrable complex structure, the metric is hermitian, there is a holomorphic 
$(3,0)$-form that  squares to the volume form, but the K\"ahler form is not
closed, and thus the metric is not K\"ahler.  In sections 6 and 7 we
show how the solutions of \refs{\PilchYG,\HalmagyiJY}  and 
\refs{\KhavaevGB,\KhavaevYG}. are contained in our far more general family.
Finally, in section 8 we make some concluding remarks.

\newsec{Some field theory constraints on the holographic dual}

The underlying field theory is ${\cal N}=4$ super-Yang-Mills
theory perturbed by a mass term for one of the three ${\cal N}=1$ adjoint chiral
superfields.  The superpotential has the form:
\eqn\gaugesuperpot{
 W ={\rm Tr} \,\big( \Phi_3
[\Phi_1,\Phi_2] \big) ~+~ \coeff{1}{2} m\, {\rm Tr}\big( \Phi_3^2\big)
\ .}
This breaks the supersymmetry to $\cN =1$, and  the theory flows to a
non-trivial ${\cal N}=1$ superconformal fixed point in the infra-red
\LeighEP.   The holographic description of the fixed point and flow
may be found in \refs{\KhavaevFB,\FreedmanGP,\PilchEJ,\PilchFU}
The fields, $\Phi_1$ and $\Phi_2$ remain massless and
there is thus a four-dimensional Coulomb branch  
described in terms of the vevs of   $\Phi_1$ and $\Phi_2$.  
A two-parameter family of holographic flows on this Coulomb branch
were studied in \refs{\KhavaevYG,\KhavaevGB}, 
and a brane-probe study can be found in \refs{\JohnsonIC,\JohnsonZE}.
For the moment we will assume that the vevs of   $\Phi_1$ and $\Phi_2$
are zero.

The ${\cal N}=4$ theory has an $SO(6)$ $\cR$-symmetry, and under the deformation
\gaugesuperpot\ this is broken to an $SU(2)$ global symmetry  and a $U(1)$ 
$\cR$-symmetry.  The $SU(2)$ acts
on $\Phi_1$ and $\Phi_2$ as a doublet, while the $\cR$ symmetry
acts on $\Phi_1$,  $\Phi_2$  and $\Phi_3$ with charges $(1/2,1/2, 1)$  \FreedmanGP:
\eqn\Rsymm{ \Phi_j \to e^{{1 \over 2}i \alpha}\, \Phi_j \,, \quad j=1,2 \,; \qquad
\Phi_3 \to e^{i \alpha}\, \Phi_3\,.}
and so both terms in the superpotential \gaugesuperpot\ have $\cR$-charge 
$2$, as they must.  If we allow the mass, $m$, to rotate by a phase then we have 
a further  $U(1)$ symmetry  under which:
\eqn\bonussymm{ \Phi_j \to e^{{1\over 2} i \alpha}\, \Phi_j \,, \quad j=1,2 \,,  \qquad
\Phi_3 \to \Phi_3 \,, \qquad m \to m \, e^{i \alpha} \,.}
This may, of course, be mixed with the $\cR$-symmetry action.

In the holographic dual,  the vevs of the  scalar fields correspond to directions 
perpendicular to the branes, and we will represent the three complex directions 
corresponding to  $\Phi_1$, $\Phi_2$ and  $\Phi_3$  by three sets of 
complex polar coordinates:  $(v, \varphi_1)$, $(w, \varphi_2)$ and $(u, \varphi_3)$.
For later convenience (and to match the conventions of earlier papers), we 
will make the field theory identifications in which, at least asymptotically:
\eqn\FTphases{\Phi_1 \sim v \, e^{ - i \varphi_1} \,, \qquad \Phi_2 \sim w \, 
e^{- i \varphi_2} \,, \qquad \Phi_3 \sim u  \, e^{+ i \varphi_3} \,.}
Thus the $SU(2)$ acts on  $(v, \varphi_1)$ and $(w, \varphi_2)$, and the 
$U(1)$  $\cR$ symmetry corresponds to:
\eqn\IRRaction{\varphi_j ~\to~ \varphi_j ~-~ \coeff{1}{2}\, \alpha\,, \quad j=1,2\,;
\qquad \varphi_3 ~\to~ \varphi_3 ~+~  \alpha\,.}
Invariance under the $SU(2) \times U(1)_\cR$ where the $U(1)_\cR$ is 
defined by \IRRaction\ means that if any of the fields depend upon the
$\varphi_j$, then they can only depend upon the sum: 
$(\varphi_1 +\varphi_2 +\varphi_3 )$.

The IIB theory also has a {\it ten-dimensional} $\cR$ symmetry 
\refs{\SchwarzQR,\SchwarzWA,\HoweSR} that  
acts on the $B$ field with charge $+1$.  The $B$ field is dual to the fermion
mass, and this symmetry corresponds to \bonussymm\ in the field theory. 
In terms of coordinates the latter symmetry is:
\eqn\coordbonus{\varphi_j ~\to~ \varphi_j - \coeff{1}{2}\, \alpha\,, \quad j=1,2\,,
\qquad \varphi_3 ~\to~ \varphi_3\,,  \qquad B \to B  \, e^{i \alpha}  \,.}
One therefore finds that the $B$ field dual to the fermion mass in
\gaugesuperpot\ must have the phase dependence:
\eqn\Bphase{ B_{\mu \nu} ~\sim~ e^{i(\varphi_1 +\varphi_2 +\varphi_3 )} \,.}
One can also deduce this result directly from linearization of the
supergravity action and using the fact that the fermion masses are dual to
the lowest modes of the tensor gauge field, $B_{\mu \nu}$.  This is also 
consistent with the results of \refs{\PilchYG, \HalmagyiJY}.

From \refs{\CorradoNV,\PilchYG, \HalmagyiJY} we also know that the family of flows we 
seek  has a complex structure that matches the intuitive complex structure provided
by the field theory.  Moreover,  by a suitable choice of the $B$-field gauge,
we may take the holographic dual of the fermion mass term to be a 
$B$ field of holomorphic type $(2,0)$.   In addition we also know that 
for these flows the dilaton background is trivial.  

In this paper we want to investigate the Coulomb branch of the flow.
We are therefore going to allow $\Phi_1$ and $\Phi_2$ to develop
vevs.   However, to keep things manageable, we are going to assume that these 
vevs are invariant under the $U(1)^2 \subset  SU(2) \times U(1)_\cR$.  That is,
the branes can spread in the $(v,w)$ directions, but will only be
allowed to do so in a manner that is independent of $(\varphi_1,  
\varphi_2)$.   Thus the metric and all the background fields will 
have a $U(1)^3$ invariance, but will be allowed to depend arbitrarily
upon $(u,v,w)$.     It is convenient to represent the  $U(1)$ symmetries in terms of
Lie derivatives.  First, there is the residual $U(1)$ subgroup of
$SU(2)$ generated by:
\eqn\residualUone{ (\cL_1 -  \cL_2 ) \,,}
where $\cL_j$ denotes the Lie Derivative  along the Killing vector defined by 
translations along $\varphi_j$.    Then there is the $\cR$ symmetry operator:
\eqn\cRoneps{\cL_\cR ~\equiv~ \cL_3- \coeff{1}{2}(\cL_1 + \cL_2 ) \,,}
Finally, there is the extra $U(1)$ is given by:
\eqn\bonusoneps{- \coeff{1}{2}\, (\cL_1 + \cL_2) ~+~  Q_{IIB} \,,}
where $Q_{IIB}$ is the IIB $\cR$ charge of the field upon which this
operator acts. 

The Killing spinors that generate the unbroken supersymmetry transformations
must transform appropriately under these $U(1)$'s.  First, before we turn on the
vevs of $\Phi_1, \Phi_2$, the supersymmetries must be a $SU(2)$ singlet, and
thus:
\eqn\firstepsDE{(\cL_1 -  \cL_2 ) \, \epsilon ~=~0 \,.}
The operator $\cL_\cR$ must generate the $\cR$-symmetry and hence:
\eqn\secondepsDE{\cL_\cR \, \epsilon ~\equiv~ (\cL_3- \coeff{1}{2}(\cL_1 + \cL_2 ) )\, 
\epsilon ~=~\Gamma^{1234}\, \epsilon\,.}
The right hand side of this equation is precisely reflects the fact
that the $\cR$-symmetry  rotates the four-dimensional spinor components with 
charges $\pm 1$ depending upon their helicity.
Under the last $U(1)$ one has:
\eqn\thirdepsDE{  \coeff{1}{2}(\cL_1 + \cL_2 )\,  \epsilon ~=~\coeff{1}{2}\, \epsilon\,,}
where we have used the fact that $Q_{IIB} = {1 \over 2}$ for the supersymmetry.

There are some signs and ambiguities in the the foregoing prescription. First, the
is the sign of the term on the right-hand side of \secondepsDE\ depends upon
spinor conventions.  Secondly, it is not obvious that the action of
\thirdepsDE\ should not be combined with a four-dimensional $\cR$
symmetry transformation of the supersymmetry parameter, but it turns out 
that the ten-dimensional chiral rotation implied by $Q_{IIB} = {1 \over 2}$
is all that one needs.   The complete justification of the foregoing
angular dependences of $\epsilon$ really comes from the fact that they
are required by the solutions of the supersymmetry conditions that we analyze
below\foot{They can also be deduced from the results of \PilchYG.}.  
Our purpose here is to make the angular behaviour of the Killing
spinors more intuitive.    Having done this, we have pinned down solution sufficiently 
to provide a  readily solvable Ansatz for the holographic dual in supergravity. 

\newsec{The Supergravity Background}

\subsec{The metric and complex structure}

We take the ten-dimensional manifold to have the usual warped product
form:  
\eqn\tenmet{ds_{10}^2 ~=~   H_0^2 (\eta_{\mu \nu}\, dx^\mu \, dx^\nu)~-~ 
H_0^{-2}\, ds_6^2 \,,}
where $ds_6^2$ is a hermitian metric on a complex manifold, 
${\cal M}_6$, transverse to the D3 branes.   Following from the field theory,  
we will parametrize this ``internal manifold'' by three complex coordinates
whose phases, $\varphi_j$, $j=1,2,3$,  generate a $U(1)^3$ symmetry of
the background. The remaining, ``radial coordinates,'' will be denoted by 
$(u,v,w)$.  Following \refs{\PilchYG,\HalmagyiJY}, it is natural to single out a complex 
coordinate  $z_3 \equiv u \, e^{i \varphi_3}$ that is to be associated with the directions
dual to the massive  chiral multiplet.  One then fibers the remaining two
complex directions over this base.   To be more explicit,  we take the complex
coordinates to be
\eqn\cplxcoords{z_1 ~\equiv~  e^{h_1 + i \varphi_1} \,, \qquad
z_2 ~\equiv~  e^{h_2 + i \varphi_2} \,, \qquad
z_3~\equiv~  u \, e^{i \varphi_3}  \,.}
for some functions, $h_j(u,v,w)$,  and introduce the holomorphic forms:
\eqn\holforms{\eqalign{\omega_1 ~\equiv~ & dh_1 + i\,d \varphi_1 -
(\del_u h_1) \,  \omega_3 ~=~ (\del_v h_1) \, dv + (\del_w h_1) \, dw + 
i\, (d \varphi_1 - u\, \del_u h_1 \, d \varphi_3) \,, \cr
\omega_2 ~\equiv~ & dh_2 + i\,d \varphi_2 -
(\del_u h_2) \,  \omega_3 ~=~ (\del_v h_2) \, dv + (\del_w h_2) \, dw + 
i\, (d \varphi_2 - u\, \del_u h_2 \, d \varphi_3) \,, \cr
\omega_3 ~\equiv~ & du + i\,u\, d \varphi_3\,.}}
We then make the metric Ansatz:
\eqn\metAnsatz{ ds_{6}^2 ~=~   A_1 \, |\omega_1|^2 
~+~ A_2 \, |\omega_2|^2  ~+~ A_3  \, |\omega_3|^2 
 ~+~  A_0 \,(\omega_1 \, \bar \omega_2 ~+~ \omega_2 \, \bar \omega_1) \,.} 
where the $A_j$ are, as yet arbitrary functions of $(u,v,w)$.
This Ansatz is a natural generalization of the results found in 
\refs{\PilchYG,\HalmagyiJY}.  The presence of the cross terms with coefficient $A_0$, 
are suggested by the angular terms noted in \refs{\KhavaevGB,\KhavaevYG}.   

The complex structure is:
\eqn\Jdefn{J~=~ A_1 \, \omega_1 \wedge \bar \omega_1 ~+~ 
A_2 \, \omega_2 \wedge \bar \omega_2 ~+~ A_3 \, \omega_3 \wedge 
\bar \omega_3 ~+~A_0 \, (\omega_1 \wedge \bar \omega_2 ~+~
\omega_2 \wedge \bar \omega_1)\,.}

An alternative way to arrive at this Ansatz is to use the reparametrization invariance 
$u \to \tilde u(u,v,w)$ to  arrange that the metric in the $(u, \varphi_3)$ directions is 
proportional to  $du^2 + u^2 d \varphi_3^2$.  We may then use the reparametrization 
invariance in $v$ and $w$ to eliminate  cross-terms of the form $du \, dv$ and $du \, dw$. 
Thus \metAnsatz\ provides the most general hermitian metric with the coordinates
fixed in this manner.    We have thus fixed the coordinates by prescribing the form
of the metric, and we  will solve for the functions that define the complex variables.
As we will see, this inversion of the usual procedure leads to a significant 
simplifications.

To define the spinors, we introduce the frames:
\eqn\frames{\eqalign{e^a ~=~ & H_0 \,  dx^a \,, \ \ \  a=1, \dots, 4 \,,  \quad  \quad 
(e^5 + i\, e^{10}) ~=~ H_0^{-1} \, H_3 \, \omega_3 \,,  \cr
\quad (e^6 + i e^9)  ~=~ & H_0^{-1} \, ( H_1 \, \omega_1 ~+~   H_4 \, \omega_2)\,, 
\qquad (e^7 + i e^8)  ~=~ H_0^{-1} \,  H_2 \, \omega_2  \,,}}
and thus:
\eqn\AHreln{A_0 ~=~ H_1\, H_4 \,, \quad A_1 ~=~ H_1^2 \,, \quad A_2 ~=~  
 (H_2^2 + H_4^2) \,,   \quad A_3 ~=~ H_3^2\,.}

\subsec{The tensor gauge fields}

Since we are dealing with a distribution of $D3$ branes, we define 
the five-form field strength in terms of a single potential function:
\eqn\fourindpot{
C_{(4)} ~=~ k \,dx^0\wedge dx^1\wedge dx^2\wedge dx^3\,, }
for some function, $k(u,v,w)$, and then take the five-form field strength to be:
\eqn\fiveform{ F_{(5)}\eql dC_{(4)}+*dC_{(4)}\,. }

The Ansatz for the two-form  potential  is simply to take the most general $(2,0)$ 
form with the appropriate phase dependence: 
\eqn\Atwopot{
B_{(2)} ~=~ - i\,e^{i(\varphi_1 + \varphi_2 + \varphi_3  )}\big[\,b_1\, \omega_2 \wedge
\omega_3 ~-~  b_2\, \omega_1 \wedge \omega_3 ~+~  b_3 \, \omega_1 \wedge
\omega_2\, \big]  \,,}
where the $b_j$ are arbitrary functions of $(u,v,w)$.  In principle 
these functions could be complex, but they turn out to be real in the
flow solution we find here.

\subsec{The supersymmetries}

It is convenient to define the supersymmetries via projection operators.
In particular it is very useful to use projectors to isolate the supersymmetries
that would be associated with $\cM_6$ were it to be a Calabi-Yau
manifold.  To that end, define the projectors:
\eqn\projzeroone{
\Pi_0  ~=~ 
{1\over 2}\big[\, \oneone -  i\,\Gamma^{1234}\,  \big]\,, \qquad
\Pi_1  ~=~  {1\over 2}\big[\, \oneone - i\,\Gamma^{78} \big]\,, }
\eqn\projtwothree{
\Pi_2  ~=~  {1\over 2}\big[\,\oneone - i\,\Gamma^{69}  \big]\,, \qquad
\Pi_3  ~=~  {1\over 2}\big[\, \oneone + i\,\Gamma^{5\,10} \big]\,.}
Define the spinor,  $\epsilon_0$, to be one that is constant and satisfies:
\eqn\projcondszero{\Pi_j \, \epsilon_0 ~=~ 0 \,, \qquad j=0,1,2,3 \,.}
One of these projections is redundant because of the helicity condition:
$\Gamma^{11} \epsilon = - \epsilon$ where $\Gamma^{11} \equiv \Gamma^{1 \dots 10}$.

Introduce the rotation matrix
\eqn\rotdef{
\cO(\beta) ~\equiv~ \cos(\coeff{1}{2} \beta )+\sin(\coeff{1}{2} \beta)\,
\Gamma^{79}*\,.}
where $*$ denotes the complex conjugation operator.  
The Killing spinor is then given explicitly by:
\eqn\killspinor{
\epsilon~=~H_0^{1\over2}\,e^{{i \over 2} (\varphi_1 + \varphi_2+ \varphi_3)}\,
\cO(\beta)\,e^{- i \varphi_3}\epsilon_0\,.}
This spinor obeys the projection conditions:
\eqn\simpprojcond{
\widehat \Pi_0  \,\epsilon ~=~ 0 \,,\qquad  \Pi_1  \,\epsilon ~=~ 0 \,,\qquad
\Pi_2  \,\epsilon ~=~ 0  \,, }
where $\widehat \Pi_0 $ is the dielectrically deformed projection operator
\refs{\PopeJP\GowdigereJF\PilchJG\PilchYG\NemeschanskyYH{--}\BenaJW}:
\eqn\nontrivproj{
\widehat\Pi_0  ~=~ 
{1\over 2}\Big[\, \oneone - i\,\Gamma^{1234}\, \big(\cos(\beta)-\,
e^{i(\varphi_1 + \varphi_2+ \varphi_3)}\sin(\beta) \,\Gamma^{7 9}*\,\big)\Big]\,, }
Also observe that the spinor, $\epsilon$, satisfies \firstepsDE, \secondepsDE\ and 
\thirdepsDE, which, in fact, determine the dependence of $\epsilon$ upon 
the angles $\varphi_j$.  The normalization factor, $H_0^{1 \over 2}$, in 
\killspinor\ is fixed by the requirement that:
\eqn\kvector{K^\mu ~=~ \bar \epsilon \,  \Gamma^\mu  \epsilon \,,}
is a Killing vector.

\newsec{Calabi-Yau Conditions}

It is very instructive to look at the the conditions on the metric \metAnsatz\
and complex structure \Jdefn\ required to make $\cM_6$ into a Calabi-Yau space.
This will not generate the flow background  that we seek, but it will come very close.

\subsec{Imposing the K\"ahler condition}

It is convenient to introduce the matrices:
\eqn\metmats{{\cal A}~\equiv~ \left(\matrix{A_1 & A_0 \cr
A_0 & A_2}\right) \,, \qquad   {\cal H}~\equiv~ \left(\matrix{v^{-1} \,
\del_v h_1 & w^{-1} \, \del_w h_1 \cr
v^{-1} \, \del_v h_2 & w^{-1} \, \del_w h_2}\right) \,,}
and set 
\eqn\matprod{{\cal B} ~\equiv~ \left(\matrix{B_1 & B_2  \cr
B_3 & B_4 }\right) ~=~    {\cal A}  \cdot {\cal H}\,.}
Then the conditions that \Jdefn\ be K\"ahler are equivalent to:
\eqn\Kcondone{ \partial_u \, {\cal B} ~=~0 \,,}
\eqn\Kcondtwo{\coeff{1}{w} \, \partial_w \, B_1   ~=~
\coeff{1}{v} \, \partial_v \, B_2 \,, \qquad  \coeff{1}{w}\, \partial_w \, B_3   ~=~
\coeff{1}{v} \, \partial_v \, B_4 \,,}
and 
\eqn\Kcondthree{ {\cal A}\cdot \left(\matrix{  u^{-1} \del_u( u \, \del_u h_1)  \cr 
 u^{-1}  \del_u( u \,\del_u h_2) }\right) ~=~ - \Big({\cal H}^{-1}\Big)^t \cdot  
\left(\matrix{  v^{-1} \, \partial_v \, A_3 \cr w^{-1}\, \partial_w \, A_3 }\right) \,,}
where the superscript $t$ denotes the transpose.

At large values of $u$ we want the metric on ${\cal M}_6$ to become 
asymptotically flat:
\eqn\asympmet{ds_6^2  \to (du^2 + u^2 d \varphi_3^2) ~+~ 
(dv^2 + v^2 d \varphi_1^2) ~+~ (dw^2 + w^2 d \varphi_2^2) \,,}
and hence, at large $u$, we must have 
\eqn\asympfns{h_1  \to \log(v)  \,, \quad h_2  \to \log(w)\,, \qquad
A_1 \to v^2  \,, \quad A_2 \to w^2 \,, \quad A_0 \to 0 \,. }
Therefore, for all values of $u$ we must have:
\eqn\BeqlOne{{\cal B} ~=~   \left(\matrix{1 & 0  \cr
0 & 1}\right)  \,,}
and thus:
\eqn\Aresult{{\cal A} ~\equiv~ \left(\matrix{A_1 & A_0 \cr
A_0 & A_2}\right) ~=~ {\cal H}^{-1} ~=~   \Delta^{-1} 
\left(\matrix{  v \, \del_w h_2 & - v \, \del_w h_1 \cr
- w \, \del_v h_2 & w  \, \del_v h_1  }\right) \,,}
where $\Delta$ is the Jacobian:
\eqn\Jac{\Delta ~\equiv~(\del_v h_1) \, (\del_w h_2) ~-~ 
(\del_v h_2)\, (\del_w h_1)  \,.}

This system of equations is elementary to analyze.  First observe 
that \Aresult\ gives $A_0, A_1$ and $A_2$  in terms of $h_j$.  Moreover,
there are two equations for $A_0$ and these imply:  
\eqn\gdefn{ v \, \del_w h_1 ~=~ w \, \del_v h_2 \qquad \Leftrightarrow  
 \qquad h_1~=~ \coeff{1}{v}\, \del_v g \,, \quad  h_2~=~ \coeff{1}{w}\, \del_w g \,,}
for some ``master function,'' g.  Finally, using $\cA = \cH^{-1}$ in \Kcondthree\
shows that:
\eqn\Athree{ A_3 ~=~ - u^{-1}\, \del_u( u\, \del_u g) \,.}
Thus the coordinates and the entire K\"ahler metric are determined
once we know the function $g$. 

\subsec{Calabi-Yau metrics}

The following is a manifestly holomorphic $(3,0)$-form on $\cM_6$:
\eqn\Omdefn{\Omega ~\equiv~ dz_1 \wedge dz_2 \wedge dz_3 ~=~
 e^{h_1 + h_2} e^{i(\varphi_1 + \varphi_2+ \varphi_3)}
\, \omega_1 \wedge \omega_2 \wedge \omega_3 \,. } 
A Ricci-flat, K\"ahler metric has the property
\eqn\ricciflat{{1 \over 3!}\,J\wedge J\wedge J=\Omega\wedge\overline\Omega\,,}
and using \Jdefn\ and \Omdefn\ this equation is equivalent to:
\eqn\flatone{ A_3\,(A_1 \, A_2 - A_0^2) ~=~ e^{2(h_1 + h_2)} \,.}
Using the K\"ahler conditions derived above, this becomes:
\eqn\CYmaster{\coeff{1}{u} \, \del_u (u \, \del_u \, g) ~+~ {\Delta \over v\, w} \,
\exp\Big( {2(\coeff{1}{v}\, \del_v g  + \coeff{1}{w}\, \del_w g)} \Big) ~=~ 0 \,.}
One can write this as a system for the $h_j$ by acting with $\del_v$ and
$\del_w$:
\eqn\CYsystem{\eqalign{{1 \over u} \, \del_u (u \, \del_u \, h_1) ~+~ & {1 \over v}\, 
\del_v\bigg( {\Delta \over v\, w} \, e^{2(h_1 + h_2)} \bigg) ~=~ 0 \cr
{1 \over u}\, \del_u (u \, \del_u \, h_2) ~+~ &{1 \over w} \, 
\del_w \bigg( {\Delta \over v\, w} \, e^{2(h_1 + h_2)} \bigg) ~=~ 0 \,,}}
which is the natural generalization of the result presented in \HalmagyiJY.

\subsec{The Killing spinors on the Calabi-Yau manifold}

Consider the spinor, $\hat \epsilon$, defined by taking $\beta =0$ and
$H_0 =1$ in \killspinor.  This spinor obeys:
\eqn\CYprojs{{1\over 2}\big[\, \oneone +  i\,\Gamma^{5\,10} \big]\, \hat \epsilon
~=~ {1\over 2}\big[\, \oneone - i\,\Gamma^{69} \big]\, \hat \epsilon
~=~{1\over 2}\big[\, \oneone - i\,\Gamma^{78} \big]\, \hat \epsilon~=~0 \,,}
and has a phase dependence:
\eqn\epshatphase{ \hat \epsilon ~=~ e^{{i \over 2} (\varphi_1 + \varphi_2 -  \varphi_3)}
\, \epsilon_0 \,.}

This spinor is, in fact, the one associated with the complex structure \Jdefn\
but with $\varphi_3 \to - \varphi_3$.  This change of complex structure is
naturally suggested by the field theory since it is consistent with the phases in
 \FTphases.    To be more explicit,  consider a new set of holomorphic forms:
\eqn\newholforms{\eqalign{\hat \omega_1 ~\equiv~ & dh_1 + i\,d \varphi_1 -
(\del_u h_1) \, \hat  \omega_3 ~=~ (\del_v h_1) \, dv + (\del_w h_1) \, dw + 
i\, (d \varphi_1 + u\, \del_u h_1 \, d \varphi_3) \,, \cr
\hat \omega_2 ~\equiv~ & dh_2 + i\,d \varphi_2 -
(\del_u h_2) \, \hat \omega_3 ~=~ (\del_v h_2) \, dv + (\del_w h_2) \, dw + 
i\, (d \varphi_2 + u\, \del_u h_2 \, d \varphi_3) \,, \cr
\hat \omega_3 ~\equiv~ & du - i\,u\, d \varphi_3 = \bar \omega_3\,,}}
 along with a complex structure defined by \Jdefn, but with 
 $\omega_i \to \hat \omega_i$.    The  spinor, $\hat \epsilon$,
 is then the Killing spinor for the Calabi-Yau metric
 associated with this complex structure  on $\cM_6$.  This change
of complex structure generates some simple sign changes in the 
analysis above.   

This observation about the complex structure will make no difference to the 
subsequent  analysis in this paper, however it will prove important if one tries to 
to interpolate between  the Calabi-Yau  flow of \refs{\KlebanovHH} and that of 
\refs{\FreedmanGP, \PilchFU} as discussed in \refs{\CorradoWX, \HalmagyiJY}.
In particular, it is important to note that the two-form tensor gauge fields
are of type $(2,0)$ with respect to the complex structure \Jdefn, but in the
$\beta \to 0$ limit  the Killing spinor of that solution is not that of the 
Calabi-Yau space based upon the complex structure \Jdefn.  Put differently,
if one starts with the Calabi-Yau flow, then the $B$-field flux is {\it not}
of type $(2,0)$ with respect to the complex structure of the Calabi-Yau
metric:  The holomorphic forms are complex conjugated in the $(u, \varphi_3)$
direction.

\newsec{The new flux solutions }

The solutions with the non-trivial background fluxes closely parallel
the Calabi-Yau solutions found in the previous section.  The metric is no longer
K\"ahler, but the two-form:
\eqn\newtwoform{\widehat J ~\equiv~   A_1 \, \omega_1 \wedge \bar \omega_1 ~+~ 
A_2 \, \omega_2 \wedge \bar \omega_2 ~+~ A_3 \, \cos(\beta) \, \omega_3 \wedge 
\bar \omega_3 ~+~A_0 \, (\omega_1 \wedge \bar \omega_2 ~+~
\omega_2 \wedge \bar \omega_1) \,,}
{\it is} closed.   In comparing this with \Jdefn, note the presence of the $\cos \beta$ 
in the third term.  This means that $\hat J$  no longer yields an almost complex
structure when combined with the metric \metAnsatz, but the closure of 
$\widehat J$ provides  a very convenient way of encoding some of 
the equations that define the new flux solutions.
This means \Kcondone, \Kcondtwo\ and \Aresult\ remain true, 
and that there is still a ``master function,''  $g$, defined by \gdefn.  Moreover,
\Kcondthree\ is replaced by:
\eqn\defKcondthree{ {\cal A}\cdot \left(\matrix{  u^{-1} \del_u( u \del_u h_1)  \cr 
u^{-1} \del_u( u \del_u h_2) }\right) ~=~  \Big({\cal H}^{-1}\Big)^t \cdot  
\left(\matrix{  v^{-1} \, \partial_v \, (A_3 \cos(\beta) ) \cr w^{-1}\, \partial_w \, 
(A_3 \cos(\beta) )  }\right) \,.}
The sign change in \defKcondthree\ as compared to \Kcondthree\ is
due to the change complex structure described in section 4.
To specify $A_3$ and $\beta$ independently, we need a further
equation, and this is:   
\eqn\newcondthree{ {\cal A}\cdot \left(\matrix{  u^{-1}  \del_u h_1   \cr 
u^{-1}  \del_u h_2} \right) ~=~ \Big({\cal H}^{-1}\Big)^t \cdot  
\left(\matrix{  v^{-1} \, \partial_v \, \big(\coeff{1}{2} A_3 \cos^2(\coeff{1}{2} \beta) \big) \cr 
w^{-1}\, \partial_w \,  \big(\coeff{1}{2} A_3 \cos^2(\coeff{1}{2} \beta) \big) ) }\right) \,.}
Combining this with \defKcondthree\ one obtains:
\eqn\Athreeeqn{ {\cal A}\cdot \left(\matrix{  u^3 \del_u( u^{-3} \del_u h_1)  \cr 
u^3\del_u( u^{-3} \del_u h_2) }\right) ~=~ -  \Big({\cal H}^{-1}\Big)^t \cdot  
\left(\matrix{  v^{-1} \, \partial_v \,  A_3  \cr w^{-1}\, \partial_w \,  A_3   }\right) \,.}

Using \Aresult\ and \gdefn, and particularly the fact that $\cA = \cH^{-1}$, one
can integrate these equations:
\eqn\intfns{  
 A_3 ~=~ - u^3 \del_u( u^{-3} \, \del_u g) ~+~ k_1(u) \,,\qquad 
 \coeff{1}{2} A_3 \cos^2(\coeff{1}{2} \beta)~=~  u^{-1}  \del_u g  ~+~  k_2(u)  \,, }
where $k_1,k_2$ are arbitrary functions of $u$.   A more detailed  
analysis of the supersymmetry variations, yields:
\eqn\uderivAthree{ A_3 ~=~ - u^3 \, \del_u \big(\coeff{1}{2} \, u^{-2}\, A_3 
\cos^2(\coeff{1}{2} \beta)  \big)\,,}
from which we obtain $k_1(u) = - u^3 \del_u( u^{-2} k_2(u))$.   To fix these functions
completely one should first note that the function $g$ is itself only defined by \gdefn\ 
up to an arbitrary function of $u$, and so we can set $k_2 = 0$ by absorbing it
into the definition of $g$.  It follows that $k_1 =0$.  Finally, note that 
at large $v, w$ we must have $A_3 \to 1$ and $\beta \to 0$, and so we
must have
\eqn\guasymp{ g(u,v,w) ~\sim~ \coeff{1}{4}\, u^2 \,, \qquad v,w \to \infty \,.}

The supersymmetry also requires \ricciflat\ and hence \flatone.  Thus we have
the following differential equation for $g$:
\eqn\master{u^3  \, \del_u (u^{-3}  \, \del_u \, g) ~+~ {\Delta \over v\, w} \,
\exp\Big( {2(\coeff{1}{v}\, \del_v g  + \coeff{1}{w}\, \del_w g)} \Big) ~=~ 0 \,,}
which can also be converted to system for the $h_j$ by acting with $\del_v$ and
$\del_w$:
\eqn\fluxsystem{\eqalign{u^3  \, \del_u \Big({1 \over u^3}\,  \del_u \, h_1\Big) ~+~ 
& {1 \over v}\,  \del_v\bigg( {\Delta \over v\, w} \, e^{2(h_1 + h_2)} \bigg) ~=~ 0 \cr
u^3  \, \del_u \Big({1 \over u^3}\,  \del_u \, h_2\Big)  ~+~ &{1 \over w} \, 
\del_w \bigg( {\Delta \over v\, w} \, e^{2(h_1 + h_2)} \bigg) ~=~ 0 \,,}}
Using \asympfns\ and \guasymp\ we find that $g$  must satisfy 
\master\ and with boundary conditions:
\eqn\gasymp{ g(u,v,w)   ~\sim~ \coeff{1}{4}\,\big[ \, u^2  ~+~ v^2 \, (2\, \log (v) - 1) ~+~  
w^2\, (2\, \log (w) - 1)\big] \,,  \quad u, v,w \to \infty\,.}
Once one finds a solution to this equation one  determines
the metric functions on $\cM_6$ via  \Aresult\ and \intfns, which we summarize
as:
\eqn\intmetfns{\eqalign{A_1 ~=~&  {v\over w\, \Delta}\, \del_w^2 g \,, \qquad 
A_2 ~=~  {w\over v\, \Delta}\, \del_v^2 g \,, \cr 
A_0 ~=~ & {1 \over \Delta}\, \del_v \del_w g\,, \qquad 
A_3 ~=~ - u^3 \del_u( u^{-3} \, \del_u g) \,.}}
The deformation angle, $\beta$, is given by \intfns, and this may be rewritten as
\eqn\betaeqn{   \cos^2(\coeff{1}{2} \beta)   ~=~  - {2\, \del_u g \over  
u^4 \del_u( u^{-3} \, \del_u g)} \,.}

The remaining parts of the solution are simple to determine from the
functions defined above.  Exactly as in \PilchYG, we have:
\eqn\Hzeroandm{H_0^2 ~=~ {a \,u \over \sqrt{A_3} \, \sin \beta} \,,
\qquad k  ~=~ - \coeff{1}{4} \, H_0^4 \, \cos\beta \,,}
where $a$ is a constant of integration, and $k$ is the function in
\fourindpot. The constant, $a$, may be absorbed into
a rescaling of the coordinates, but it is often convenient to retain it.
Finally, the two-form flux functions, $b_j$, are given by:
\eqn\bjsolns{ b_1 ~=~ \coeff{2}{a\, u} \, e^{h_1 + h_2} \, \del_u h_1\,, \qquad  
b_2 ~=~ \coeff{2}{a\, u} \, e^{h_1 + h_2} \, \del_u h_2\,, \qquad
b_3 ~=~ - \coeff{2}{a\, u} \, \sin^2(\coeff{1}{2} \beta) \, e^{h_1 + h_2}  \,.}
One can re-write the fluxes in a slightly more compact form using \cplxcoords\
and \Atwopot:
\eqn\Atwopot{
B_{(2)} ~=~ { 2 i\over \bar z_3} \, \big[\,  dz_1 \wedge dz_2  ~-~ z_1\, z_2 \,
\cos^2(\coeff{1}{2} \beta) \, \omega_1 \wedge \omega_2\, \big]  \,.}
It is also worth noting that if we convert this to frames then $B_{(2)} $
has the following component in the $6789$ direction:
\eqn\Atwopotproj{
a \, i\,\tan(\coeff{1}{2} \beta)\, e^{i(\varphi_1 + \varphi_2 + \varphi_3  )}\,(e^6 + i e^9)  \wedge
(e^7 + i e^8)\,,}
exactly as in \PilchYG.

Thus,  we once again find a family of solutions that are almost Calabi-Yau
in that the metric is hermitian with respect to an integrable complex structure, and
the holomorphic $(3,0)$ form satisfies \ricciflat.  The metric is simply not
K\"ahler, and there is a non-trivial flux that can be arranged to have a
potential of $(2,0)$ type.   The underlying master differential equation that governs 
our new solution is a relatively simple deformation of the equation that governs
the corresponding Calabi-Yau  metrics, and follows a very similar pattern to that
obtained in \PilchYG.

\newsec{Radial Coulomb branch flows}

The geometry of $\cN=1$ supersymmetric Coulomb branch flows  in which the
branes are allowed to spread  purely {\it radially} were analyzed in 
\refs{\PilchYG,\HalmagyiJY}.   These flows preserve  the $SU(2)$ global symmetry 
that rotates $(\Phi_1, \Phi_2)$ as a doublet.  Accordingly, the complex coordinates
analogous to \FTphases\ were taken to be:
\eqn\oldcoords{\Phi_1 \sim V \,  \cos(\coeff{1}{2} \phi_1)  \, e^{- {i \over 2}\,
 (\phi_2 + \phi_3)} \,, \qquad \Phi_2 \sim V \, \sin(\coeff{1}{2} \phi_1) \, 
  e^{{i \over 2}\,  (\phi_2 - \phi_3)}\,,  \qquad \Phi_3 \sim u  \, e^{ i \phi}  \,.}
By comparing this to \FTphases\ we see that the relevant change of coordinates is:
\eqn\coordtrf{
v ~=~ V  \cos(\coeff{1}{2} \, \phi_1)  \, \qquad w = V \sin(\coeff{1}{2} \,\phi_1)\,, \qquad
\varphi_1 = \coeff{1}{2}(\phi_3 + \phi_2)\,, \qquad  
\varphi_2 = \coeff{1}{2}(\phi_3 - \phi_2)\,.}
Similarly, the analog of \cplxcoords\ was:
\eqn\oldcoords{
z_1 ~=~ e^{{1\over 2}\,  \Psi } \,  \cos(\coeff{1}{2} \phi_1)  \,e^{{i \over 2}\,
(\phi_2 + \phi_3)} \,, \quad z_2 ~=~ e^{{1\over 2}\,
 \Psi } \, \sin(\coeff{1}{2} \phi_1)  \,  e^{-{i \over 2}\,  (\phi_2 - \phi_3)}\,, 
 \qquad z_3 ~=~ u\, e^{- i\, \phi}  \,. }
and hence one has:
\eqn\hPsireln{ h_1 = \coeff{1}{2}\, \Psi ~+~ \log \cos (\coeff{1}{2} \, \phi_1) \, \qquad
 h_2 = \coeff{1}{2}\, \Psi  ~+~  \log \sin (\coeff{1}{2}\, \phi_1) \,.}
One can then easily show that:
\eqn\deltares{ {\Delta \over v \, w} \, e^{2(h_1 + h_2)} ~=~   {e^{2 \Psi} \over 2 V^3} \,
{\partial \Psi \over \partial V}  \,,}
and from this it follows that both equations in \CYsystem\ reduce to:
\eqn\oldCYmaster{\coeff{1}{ u} \,  \partial_u \big( u \, \partial_u \Psi \big) ~+~  \coeff{1}{V}  \,
\partial_V \big( \coeff{1}{V^3} \, e^{2 \Psi}\, \partial_V \Psi   \big) ~=~ 0\,,}
while both equations in \fluxsystem\ reduce to:
\eqn\oldmaster{u^3  \,  \partial_u \big( \coeff{1}{ u^3} \, \partial_u \Psi \big) ~+~  \coeff{1}{V}  \,
\partial_V \big( \coeff{1}{V^3} \, e^{2 \Psi}\, \partial_V \Psi   \big) ~=~ 0\,.}
This exactly reproduces the results of \PilchYG.

One should also note that to completely solve the flow solution in \PilchYG\ it required an
auxiliary function defined by:
\eqn\Sdefn{ 
{\partial \cS \over\partial u} ~=~  -{1\over 2 \,u^3\, V^3 }\,{\partial e^{2\Psi}\over\partial V}\,,
\qquad {\partial \cS \over \partial V} ~=~  {V \over u^3}\,{\partial \Psi\over\partial u} \,.}
If one uses \deltares, \master\ and \gdefn\ then one finds that:
\eqn\Seqn{ \cS ~=~ \coeff{2}{u^3} \, \partial_u g \,.}
In other words, in \PilchYG\ one could have viewed the second equation in 
\Sdefn\ as implying the existence of a function, $g$, with:
\eqn\geqn{ \cS ~=~ \coeff{2}{u^3} \, \partial_u g \,, \qquad
\Psi  ~=~ \coeff{2}{V} \, \partial_V g\,, }
and then the first equation in \Sdefn\ becomes a differential equation
for $g$, and this is simply the reduction of \master.  Thus, by recasting 
the entire problem in terms of $g$, one sees that there is nothing in the work
of \PilchYG\ that is special to two variables:  All the interesting functions merely
emerge as partial derivatives of the single, underlying master function, $g$.

\newsec{Gauged supergravity flows}

There is a three-parameter family of solutions that should be among
the new flux solutions presented in section 5.  This family was obtained 
by ``lifting'' solutions of  five-dimensional, gauged supergravity 
\refs{\KhavaevGB,\KhavaevYG}, and we now 
show how it fits into our more general class of solutions.   We will first
summarize the details of the original gauged supergravity solution and 
identify its integrable complex
structure by giving the holomorphic $(1,0)$-forms.   We then re-write the
metric in terms of these forms as in \metAnsatz\ and identify the functions,
$A_0,\dots, A_3$.  Having done this we can read off the master functions, 
$h_1$ and $h_2$, and then verify that they determine all the other functions
in the solution, as outlined in section 5.  In particular, one can verify that 
that the equations of motion in five-dimensional supergravity imply 
that the master equations, \fluxsystem, are satisfied.

\subsec{The known solution and its holomorphic structure}

In five-dimensions this solution is characterized in terms of three scalar fields,
denoted by $\chi, \nu$ and $\rho$,  and a superpotential, $W$.
The superpotential is \refs{\KhavaevGB,\KhavaevYG}:
\eqn\Wpot{
W ~=~ {1 \over 4 } \, \rho^4   \, \big(\cosh(2\chi) - 3 \big)   ~-~  {1 \over 4 \,\rho^2 } \,
(\nu^2 + \nu^{-2} ) \, \big( \cosh(2\chi) +1 \big) \,  ,}
and the equations of motion are:
\eqn\floweqs{ {d \rho \over d r} \eql  {1 \over 6 L}\, \rho^2 \, {\del W \over \del \rho}  \,, 
\qquad {d \nu \over d r} ~=~  {1 \over 2 L}\,  \nu^2  \, {\del W \over \del \nu} \,,\qquad 
{d \chi \over d r} \eql   {1 \over L}\, {\del W \over \del \chi}  \ .} 
The five-dimensional metric is then given by
\eqn\fivemet{ds_{1,4}^2 ~=~ dr^2 ~+~ e^{2 A(r)}\big(
 \, \eta_{\mu \nu} \, dx^\mu \, d x^\nu \big) \,,}
where
\eqn\Aeqn{{d A \over d r} ~=~  - {2 \over 3\,L}\,W \,.}

To lift this to ten dimensions the authors of \refs{\KhavaevGB,\KhavaevYG} introduced 
the following coordinates to parametrize the five-sphere in $\IC^3$:
\eqn\varchng{\eqalign{ z_1 ~\equiv~& x_1+ i \, x_2 ~=~
\cos \theta \,\cos \phi \, e^{i\, \varphi_1} \,,  \quad
z_2 ~\equiv~ x_3 - i \, x_4 ~=~
\cos \theta \,\sin \phi \, e^{- i\, \varphi_2}\,, \cr
z_3 ~\equiv~& x_5 - i \, x_6 ~=~
\sin \theta  \, e^{- i\, \varphi_3} \,.}}
The five-dimensional solution then lifts to the ten-dimensional metric:
\eqn\tenmetric{ds_{10}^2 ~=~ \Omega^2\,  ds_{1,4}^2 ~+~ ds_5^2\,,}
where $\Omega$, is given by:
\eqn\warpfac{\Omega  ~\equiv~ 
(\cosh\chi)^\half \,\big( \rho^{-2} \,(\nu^2 \cos^2 \phi + \nu^{-2} \sin^2 \phi) \,
\cos^2 \theta ~+~ \rho^4 \, \sin^2 \theta\big)^{1 \over 4}\,.} 
The metric, $ds_5^2$, is a complicated metric on the deformed five-sphere:
\eqn\defSfive{\eqalign{ds_5^2 ~=~ & L^2\, \Omega ^{-2}\,
\Big[\rho^{-4} \, \big(\cos^2 \theta + \rho^6\, \sin^2 \theta \,
(\nu^{-2}   \cos^2\phi  +\nu^2\, \sin^2\phi) \big)\, d\theta^2 \cr
& + \rho^2 \cos^2 \theta \,(\nu^2 \,
\cos^2\phi  +\nu^{-2} \sin^2\phi) \, d\phi^2  \cr
&-2\,\rho^2 \, (\nu^2   - \nu^{-2} ) \,\sin\theta\, \cos \theta\,
\sin\phi\,\cos \phi\, d\theta \,d \phi \cr
&+~  \rho^2\, \cos^2 \theta \, (\nu^{-2}  
\cos^2\phi \, d\varphi_1^2  + \nu^2\, \sin^2\phi\, \, d\varphi_2^2) 
~+~ \rho^{-4} \, \sin^2 \theta \,  d\varphi_3^2 \,\Big]\cr
& +  L^2\,\Omega^{-6}\,\sinh^2 \chi\, \cosh^2 \chi \,\big(\cos^2\theta
\, (\cos^2 \phi \,  d\varphi_1 - \sin^2 \phi \,  d\varphi_2)  - 
\sin^2 \theta \,  d\varphi_3\,\big)^2 \,.}}
where $L$ is the radius of the round sphere.  

To relate this to the solution presented here one makes the change of variable:
\eqn\uvwchange{u ~=~  e^{{3 \over 2} \, A} \sqrt {\sinh \chi}  \sin \theta \,,   \quad 
v   ~=~    e^A \rho \, \nu^{-1} \cos \theta \cos \phi \,, \quad  w   ~=~   
e^A \rho \, \nu  \cos \theta \sin \phi  \,.}
The metric \defSfive\ can  now be written in terms of the 
(integrable) holomorphic forms:
\eqn\holformsres{\eqalign{
\omega _3 & \equiv   du ~-~ i\, u\, d \varphi_3  \,, \cr 
\omega_1 & \equiv    dv ~-~  v\, \mu_1 ~+~  i\, v\, d\varphi_1 ~+~ 
{v \over u}\, \Big( {\nu^2 \sinh^2\chi\,  \sin^2 \theta  \over   X_0}\Big) \, \omega_3  \,, \cr
\omega_2 & \equiv   dw ~-~ w\,\mu_2 ~+~  i\, w\,d \varphi_2  ~+~
{w \over u} \, \Big({ \nu^{-2} \sinh^2\chi\, \sin^2 \theta  \over  X_0}\Big)\, \omega_3 \,,}}
where
\eqn\functiondefsa{\eqalign{\mu_1 & ~\equiv~   {\nu^2 s^2 \over L \rho^2} d r\,, 
\qquad  \mu_2 ~\equiv~  {\nu^{-2} s^2\over L \rho^2} d r \,, \cr
X_0 & \equiv   \rho^6 \sin^2 \theta  ~+~ c^2\, \cos^2 \theta \,
\big( \nu^2 \cos^2 \phi ~+~ \nu^{-2} \sin^2 \phi \big)  \,,}}
and we have adopted the convenient shorthand:
\eqn\csdefn{ c ~\equiv~ \cosh\chi \,, \qquad s ~\equiv~ \sinh\chi \,.}
The six-dimensional metric metric that underlies \tenmetric\ is:
\eqn\sixmetric{ds_6^2 ~=~    L^{-2}\, e^{2 \, A} \, \big(\Omega^4\, dr^2 ~+~ 
\Omega^2\, ds_5^2\big) \,,}
has precisely the form \metAnsatz\ with
\eqn\Acoeffs{\eqalign{
A_1 ~=~ & Y_0^{-1} (\rho^6 \sin^2 \theta +  \cos^2 \theta (c^2\, \nu^2 \cos^2 \phi +  
\nu^{-2} \sin^2 \phi)) \,, \cr
A_2 ~=~ & Y_0^{-1} (\rho^6 \sin^2 \theta +  \cos^2 \theta (\nu^2 \cos^2 \phi +  
c^2\,  \nu^{-2} \sin^2 \phi) )\,, \cr
A_3 ~=~ &  {e^{2 \, A}\, c^2\, \sin^2 \theta \over \rho^4 \, u^2 \, X_0} \,  Y_0 \,, \qquad
A_0 ~=~ {e^{- 2 \, A} \, s^2 \,v \, w \over \rho^2\, Y_0}  \,,}}
where
\eqn\Yzerodefn{ Y_0~\equiv~   \rho^6 \sin^2 \theta +  \cos^2 \theta\,  (\nu^2 \cos^2 \phi +  
\nu^{-2} \sin^2 \phi)  \,.}
To verify this one first uses the co-ordinate transformations in \uvwchange\ to
obtain the holomorphic forms, $\omega_i$, of \holformsres\ in terms of the
co-ordinates ($r, \theta, \phi$).  One can then use these expressions along with  
metric functions, $A_i$, given in \Acoeffs\ to obtain,  after a rather long but
straightforward computation, the metric in \defSfive.

\subsec{The master functions}

By comparing \newholforms\ and \holformsres, we can read off the exterior
derivatives of the functions, $h_j$:
\eqn\hjpartials{d\, h_1 ~=~ {dv \over v} ~-~ \mu_1 \,, \qquad 
d\, h_2 ~=~ {dw \over w} ~-~ \mu_2\,.}
One also finds the conditions:
\eqn\upartials{\partial_u h_1 ~=~ \Big( {\nu^2 \sinh^2\chi\,  \sin^2 \theta  \over u\,  
X_0}\Big) \,, \qquad \partial_u h_2 ~=~  \Big({ \nu^{-2} \sinh^2\chi\, \sin^2 \theta 
 \over  u\, X_0}\Big)\,.}
One can easily check that this is consistent with \hjpartials, indeed, using
the change of variables \uvwchange\ one can check that:
\eqn\muuvw{\eqalign{\mu_1 ~=~ {e^{- 2 \, A} \, s^2   \over \rho^2\, X_0}  \big( \nu^4\, 
v dv ~+~ w dw\big) ~+~  {\nu^2\, s^2 \,  \sin^2 \theta  \over u\,   X_0} \, du  \,, \cr
\mu_2 ~=~  {e^{- 2 \, A} \, s^2   \over \rho^2\, X_0}  \big(  v dv ~+~ 
\nu^{-4}\, w dw\big) ~+~  { s^2 \,  \sin^2 \theta  \over \nu^2 \, u\,   X_0} \, du  \,.}}

Define functions:
\eqn\qdefns{ q_1  ~\equiv~  \int \,  {\nu^2 s^2 \over L \rho^2} \,  d r\,, 
\qquad q_2  ~\equiv~  \int \,   {\nu^{-2} s^2\over L \rho^2}  \, d r \,,}
 and then we have
\eqn\hjres{h_1 ~=~ \log(v) ~-~ q_1 \,, \qquad  h_2 ~=~ \log(w) ~-~ q_2 \,.}
While we do not know explicit expressions for these function individually 
one can easily show that:
\eqn\exphh{ e^{2\,(h_1 + h_2)} ~=~ v^2 \, w^2\,  e^{-A}\, \rho^{-4} \,c^2 \, s^{-1} \,,}
which greatly simplifies \fluxsystem.

Note that apart from the trivial log terms, the functions, $h_j$ are
functions of only the original radial coordinate in anti-de Sitter space.
We therefore have the $h_j$ implicitly in terms of $u,v$ and $w$.  One
can use the equations of motion \floweqs\ to verify that $h_1$ and $h_2$
satisfy \fluxsystem.

One can now obtain the function, $g$, by integrating:
\eqn\gpartials{h_1 ~=~ {1 \over v}\, {\partial g \over \partial v} \,, \qquad
h_2 ~=~ {1 \over w}\, {\partial g \over \partial w} \,, \qquad
A_3 ~=~   -u^3 \,{\partial  \over \partial u}  \Big( {1 \over u^3}\, {\partial g \over 
\partial u} \Big) \,.}
It is trivial to integrate the log terms to obtain:
\eqn\ggtilde{  g~=~    \coeff{1}{4} \, (v^2~+~ w^2) ~+~ \coeff{1}{2} \,v^2
\, \log(v)  ~+~ \coeff{1}{2} \,w^2 \, \log(w) ~+~ \tilde g \,,}
where
\eqn\ggtildederivs{\coeff{1}{v}\,\partial_ v \tilde g   ~=~    - q_1  \,,
\qquad  \coeff{1}{w}\,\partial_ w \tilde g   ~=~    - q_2 \,, \qquad
A_3 ~=~   -u^3 \,{\partial  \over \partial u}  \Big( {1 \over u^3}\, {\partial 
\tilde g \over  \partial u} \Big)\,.}
Since the $q_j$ are only known explicitly as functions of $r$, and thus
implicitly as a function of $(u,v,w)$, one can, at best hope to determine
$\tilde g$ as a function of $r, \theta$ and $\phi$.  One can make significant progress
in doing this explicitly, and in this is described in the Appendix.  We have shown
here exactly how the solution of \refs{\KhavaevGB,\KhavaevYG} appears as a special 
solution to the general class of solutions derived in section 5.

\newsec{Conclusions}

We have defined a large class of $\cN=1$ supersymmetric holographic flow
solutions, and reduced them to finding a ``master function'' that is the
solution of a single partial differential equation.   As has been observed in other
papers \refs{\GowdigereJF\PilchJG\PilchYG{--}\NemeschanskyYH}, such differential 
equations naturally linearize at infinity and have a 
straightforward perturbation expansion.  More significantly, this equation is
once again a rather simple deformation of the Calabi-Yau condition.  Indeed, the
entire geometry is, once again, almost Calabi-Yau in that it has an integrable
complex structure with a hermitian metric.  There is also a holomorphic $(3,0)$ form, 
$\Omega$, such that $\Omega \wedge \bar \Omega$ is the volume form.  
The crucial difference between our solution and a Calabi-Yau compactification 
is that the metric is not K\"ahler, and there is a non-trivial, non-normalizable 
$3$-form flux.   This flux dielectrically polarizes the $D3$-branes into $D5$-branes
\MyersPS\  and this is reflected in the deformation of one of the projectors that defines 
the supersymmetry 
\refs{\PopeJP\GowdigereJF\PilchJG\PilchYG\NemeschanskyYH{--}\BenaJW}.

Our results here represent a significant extension of the results of \PilchYG.  On
a technical level we have found a class of solutions with a smaller amount of 
symmetry, in which the initial Ansatz is based upon functions of three variables.
On a more physical level, we have a family of flows that probes {\it two} independent
directions of the Coulomb  branch of the non-trivial $\cN=1$ supersymmetric 
fixed-point.   That is, our solutions describe brane configurations that can spread 
in independent radial distributions in each of the two complex directions of the 
Coulomb branch of the $\cN=1$ supersymmetric flows.  We have shown

While we have completely characterized our solutions via a simple
deformation of the Calabi-Yau conditions, this deformation is expressed
rather technically in terms of changing coefficients of the master 
differential equation.  This must have some more natural geometric 
interpretation.  The results presented here highlight the strong connections
to Calabi-Yau geometry, and as a result will provide a very useful basis for
investigating the deformation of the geometry.  Work on this is proceeding.

 \bigskip
\leftline{\bf Acknowledgements}

This work was supported in part by funds
provided by the U.S.\ Department of Energy under grant number 
DE-FG03-84ER-40168.    We would also like to thank N.~Halmagyi,
D.~Nemeschansky and K.~Pilch for helpful conversations.
 
\vfill\eject
\appendix{A}{Finding the master function of the gauged supergravity flow}

One can make significant progress in partially integrating  \ggtildederivs\
to obtain $\tilde g(r,\theta,\phi)$.  The first step is to write these equations in
terms of derivatives with respect to $r,\theta$ and $\phi$.  The first two equations
in \ggtildederivs\ reduce to:
\eqn\newgderivsa{ 
 \partial_\phi \tilde g ~= ~   e^{2\, A} \, \rho^2 \,  ( q_1\, \nu^{-2}  - q_2 \, \nu^2)\,
 \cos^2 \theta \, \sin \phi \, \cos \phi \,, }
\eqn\newgderivsb{ \eqalign{
(\partial_r - (\partial_r \log(u)) \, \tan \theta\,  \partial_\theta)\, \tilde g ~= ~  
- L^{-1}\, e^{2\, A} \, \Big[  & c^2\, (q_1 \, \cos^2 \phi ~+~q_2 \, \sin^2 \phi) \,  
\cos^2 \theta  \cr &  ~+~  \rho^6\, (\nu^{-2}\, q_1 \, \cos^2 \phi ~+~\nu^2 \,  
q_2 \, \sin^2 \phi) \,   \sin^2 \theta \Big] \,.}}
where $u(r,\theta)$ is defined by \uvwchange. 
Observe that the differential operator on the left-hand side of \newgderivsb\ 
annihilates the coordinate, $u$, and so these two equations indeed
only give information about the $v$ and $w$ dependence of $\tilde g$.

Equation \newgderivsa\ is trivial to integrate and yields
\eqn\gstepone{\tilde g = -\coeff{1}{4}\,  e^{2\, A} \, \rho^2 \,  ( q_1\, \nu^{-2}  - 
q_2 \, \nu^2)\,  \cos^2 \theta \,  \cos (2 \, \phi) ~+~  p (r, \theta) \,,}
for some function, $p (r, \theta)$. One can now substitute this into 
 \newgderivsb\  to obtain an equation for $p (r, \theta)$.  
 
At this point it is convenient to introduce a change of variables:
\eqn\yetnewvars{\eqalign{
z ~\equiv~ & \coeff{1}{2}\, \log (u) ~=~  \coeff{1}{4}\, \log \Big(e^{3\, A} \sinh \chi \,
\sin^2 \theta \Big) \,, \cr t ~\equiv~  & \coeff{1}{2}\, \log \Big({u \over \sin^2 \theta}
\Big) ~=~   \coeff{1}{4}\, \log \Big({e^{3\, A} \sinh \chi \over \sin^2 \theta}  \Big) \,.}}
We then find:
\eqn\pderiv{
\partial_t \,p ~=~ -\coeff{1}{2}\, e^{2\, A} \, \rho^{-4} \, \Big[  c^2\, (q_1 +  q_2 ) \,  
(1 -\sin^2 \theta)   ~+~  \rho^6\, (\nu^{-2}\, q_1 + \nu^2 \, q_2 )  \, \sin^2 \theta \Big] \,.}
Note that $\sin  \theta = e^{z-t}$ while  $e^{z+t}$ is purely a function of
$r$, and so  \pderiv\ has the form:
\eqn\pderivform{\partial_t \,p ~=~ f_1(z+t) ~+~ e^{4 z} \, f_2(z+t) \,,}
for some functions, $f_1$ and $f_2$.  This is trivially solved by quadrature,
and the result is:
\eqn\presult{\eqalign{
p(r,\theta)~=~   r(u) & ~-~  \coeff{1}{2\, L} \, \int \, dr\, e^{2 A}\, c^2\, (q_1 + q_2)  \cr
& ~+~  \coeff{1}{2\, L}\, u^2\, \int \, dr \, e^{-A} \, s^{-1}  \, \big(   c^2 \, (q_1 + q_2) ~-~  
 \rho^6 \,   ( \nu^{-2}\,  q_1 + \nu^2 \,  q_2) \big)  \,.}}
where $r(u)$ is some, as yet, arbitrary function of $u$, and we have used the 
fact that $e^{4 z}= u^2$.  The function, $r(u)$, can then be determined by 
substituting  \gstepone\ and \presult\ into the third equation in \ggtildederivs.

\vfill\eject
\listrefs
\vfill\eject
\end